\documentclass[aps,prd,superscriptaddress,
twocolumn,
longbibliography,
nofootinbib,floatfix,preprintnumbers,eprint]{revtex4-2}
\usepackage[utf8]{inputenc}
\usepackage{amsmath}
\usepackage{amsfonts}
\usepackage{amsthm}
\usepackage{amssymb}
\usepackage{float}
\usepackage{braket}
\usepackage{graphicx}
\usepackage{xcolor}
\usepackage{url}
\usepackage[colorlinks=true, pdfstartview=FitV, linkcolor=red, citecolor=blue, urlcolor=blue]{hyperref}
\usepackage{slashed}
\usepackage[normalem]{ulem}

\usepackage[capitalise]{cleveref}
\usepackage{placeins}

\begin{document}
\title{Looking at the Entropy in a Proton through a QGP Lens}

\preprint{MIT-CTP/6037}

\author{\mbox{Dmitri E. Kharzeev}}
\email[]{dmitri.kharzeev@stonybrook.edu}
\affiliation{Center for Nuclear Theory, Department of Physics and Astronomy, Stony Brook University, Stony Brook, New York 11794-3800, USA}
\affiliation{Energy and Photon Sciences Directorate, Condensed Matter and Materials Sciences Division,
Brookhaven National Laboratory, Upton, New York 11973-5000, USA}

\author{Krishna Rajagopal}
\email[]{krishna@mit.edu}
\affiliation{MIT Center for Theoretical Physics --- a Leinweber Institute, Massachusetts Institute of Technology, Cambridge MA 02139, USA}

\bibliographystyle{unsrt}

\begin{abstract}
We investigate the interplay between the thermodynamic (Gibbs) entropy of quark–gluon plasma (QGP) and the quantum entanglement entropy characteristic of confined hadronic states across the quark–hadron phase transition. In the deconfined regime, entropy is well described by the statistical mechanics of colored quarks and gluons.
Upon hadronization, however, the macroscopic Gibbs entropy of the plasma cannot simply vanish; instead, it is reorganized into 
the configurational entropy of a gas of colorless hadrons
together with 
quantum correlations among the confined partons within each hadron. We show that the entanglement entropy of the internal partonic wave functions inside hadrons provides a natural repository for this ``converted'' thermodynamic entropy, reconciling the apparent reduction of macroscopic entropy 
with the second law of thermodynamics. Either by extrapolating from known facts about deep inelastic scattering, or starting from  model descriptions of the proton wave function, or starting from the Hagedorn spectrum of its resonances, we provide three estimates of the magnitude of the entanglement entropy carried by a proton,
with very different uncertainties.
All three estimates indicate that
the internal entanglement entropy of the proton is similar in magnitude to the Gibbs entropy of the QGP droplet from which the proton formed as QGP cools through the 
quark-hadron transition, as for example throughout the universe microseconds after the Big Bang.
These results suggest that entanglement entropy offers a bridge between the quantum information content of hadronic states and the thermodynamic entropy of the quark–gluon plasma, providing a new lens on the microscopic mechanism of confinement and the nature of the QCD phase transition.
\end{abstract}

\maketitle

\section{Introduction}
\label{sec:introduction}

The first protons in the Universe formed from hot quark-gluon plasma (QGP) around ten microseconds after the Big Bang as the universe cooled through the QCD crossover
temperature $T_c\sim (155-160)$~MeV and the strongly coupled liquid QGP that filled the universe fell apart into a mist of hadrons. 
In this poetic sense, protons are born from QGP.  But, does it make sense to think of protons as being made of QGP?  Can protons be thought of as (very small) droplets of the primordial QGP from which they first formed? 
The energy density of a proton is comparable to that of QGP at or just above $T_c$ and, as we shall discuss in Section~\ref{sec:thermodynamics}, lattice QCD calculations~\cite{Shanahan:2018pib,Shanahan:2018nnv,Hackett:2023rif} and electron-scattering data~\cite{Polyakov:2002yz,Burkert:2018bqq,Lorce:2018egm,Polyakov:2018zvc} suggest that to the extent that the proton has an interior pressure it is comparable to the pressure of QGP at a temperature 
that we shall call $T_p$, with $T_p\gtrsim T_c$.
Nevertheless, most would say that protons can be thought of as droplets of QGP only in a poetic sense because QGP is hot, with significant thermodynamic entropy, whereas protons are zero temperature pure states, eigenstates of the Hamiltonian of QCD.  

These considerations can, however, be turned on their head. If we take the poetic view seriously, we must ask about the apparent ``entropy deficit''.  Since protons formed from droplets of a high entropy liquid, where has the thermal entropy that they inherited from the hot QGP gone?
How is the thermodynamic Gibbs entropy of the QGP from which the first protons formed reconciled with the quantum properties of confined hadrons, while respecting the second law of thermodynamics?
In this paper, we propose that the missing entropy is reorganized into \emph{entanglement entropy} carried by the partonic wave functions inside hadrons. This perspective provides a natural link between QCD thermodynamics and quantum information in QCD.

After some further introduction to the conceptual basis for our considerations in Section~\ref{sec:conceptual-basis}, we begin in Section~\ref{sec:thermodynamics} by taking the poetic question seriously and asking how much thermodynamic entropy each proton inherits from the QGP of which it formed. We find that, via this logic, each proton should have an entropy $S_{\rm thermal}\sim (5-8) k_B$. 
How does this compare to the entanglement entropy between one parton in a proton wave function and the rest of the partons in the proton? 
In previous literature~\cite{Hentschinski:2024gaa,Kharzeev:2021yyf},
the longitudinal component of the entanglement entropy has been estimated, and is $\sim 2~k_B$ for partons with large longitudinal momentum fraction $x$ 
whereas it is
$\gtrsim 3~k_B$
when one boosts to high rapidity and includes partons with smaller $x$.
Since the complete entropy should reflect our ignorance of the transverse position and momentum of each parton as well as of their longitudinal position and momentum, and since there is no transverse analogue of boosting to high rapidity, the simplest estimate for the transverse entanglement entropy is that it should be roughly twice the large-$x$ longitudinal entropy.  Adding the longitudinal and transverse contributions yields the crude estimate that the entanglement entropy of a parton in the proton wave function is $S^{(p)}_{\rm entanglement}\gtrsim 7 k_B$.  

In 
Section~\ref{sec:entanglement}, 
we shall come at this from a different starting point, estimating the entanglement entropy inside a single proton using 
basic known facts about the internal structure of the proton in QCD, 
finding $S^{(p)}_{\rm entanglement}\gtrsim (7-8) k_B$ --- consistent with the crude estimate.

Then, in Sections~\ref{sec:entanglement-entropy-via-Hagedorn} and  \ref{sec:internal-temperature},
we shall come at this from yet a third starting point that can be thought of as 
picturing the internal entropy of a proton as being in a certain sense both thermal and entanglement entropy. 
We start by taking seriously the notion that the proton has an 
effective internal temperature,
which is to say that its reduced density matrix after tracing over unobserved degrees of freedom can be approximated as thermal, and 
use this as well as what is known about the Hagedorn density of
resonance excitations of the proton to estimate its internal entanglement entropy.
We show in 
Sections~\ref{sec:entanglement-entropy-via-Hagedorn} and \ref{sec:internal-temperature} that
positing such a perspective 
leads us to an alternative estimate of the entanglement entropy within a proton, 
$S^{(p)}_{\rm entanglement}\sim (5-9) k_B$.

The three estimates that we have made of the entanglement entropy $S^{(p)}_{\rm entanglement}$ of protons today are all qualitative, but the origins of the dominant uncertainties in each estimate are completely different. And, all three estimates are
strikingly similar in magnitude to each other and to the thermal entropy $S_{\rm thermal}$ of the droplet of QGP from which protons formed microseconds after the Big Bang.
This suggests that there is no entropy deficit at all: entropy is in fact a common thread across the QCD transition, although the way in which it is woven into the fabric of the universe changes.
With this perspective, entropy from the hot Big Bang is converted into entanglement entropy within protons (and other hadrons) as they form at the QCD phase transition. Billions of years later, when experimentalists collide nuclei at the Relativistic Heavy Ion Collider (RHIC) or the Large Hadron Collider (LHC), they are liberating the entanglement entropy within the incident protons and neutrons --- as well as creating additional entropy --- all of which becomes the thermodynamic entropy of the droplet of hot QGP created in a heavy ion collision.
Or, when an experimentalist does a deep inelastic scattering experiment by measuring the scattered electron and gaining information about the longitudinal momentum of the quark struck by an incident electron, they are liberating a part of the 
entropy within the proton while remaining ignorant of the information encoded in transverse correlations among the partons
in the wave function of the proton.  We close in Section~\ref{sec:conclusions} by speculating 
how this view of the QCD history of the 
universe  ---  with thermodynamic entropy from hot QGP converted into entanglement entropy of 
hadronic wave functions at the QCD phase
transition and later released in collisions ---  
can yield further insights into the nature of confinement and the QCD phase transition and  motivate measurements at the future Electron-Ion Collider (EIC) that can shed light on both.

\section{Entanglement, Resolution Scale, and Emergent Thermality of the Proton}
\label{sec:conceptual-basis}

The proton in isolation is described by a pure quantum state $|p\rangle$, which is an eigenstate of the full QCD Hamiltonian. As such, the corresponding density matrix
\begin{equation}
\rho = |p\rangle \langle p|
\end{equation}
has vanishing von Neumann entropy. This statement, however, refers to the complete Hilbert space of QCD, including all momentum modes. In practice, both theoretically and experimentally, one never has access to this full set of degrees of freedom. Instead, physical observables are defined at a finite resolution scale, which introduces an effective ultraviolet cutoff $\Lambda$ on the accessible modes.

The introduction of this scale induces a decomposition of the Hilbert space into resolved (infrared) and unresolved (ultraviolet) sectors,
\begin{equation}
\mathcal{H} = \mathcal{H}_{\rm IR} \otimes \mathcal{H}_{\rm UV},
\end{equation}
where the cutoff $\Lambda$ defines the boundary between the two. Observables at a given resolution scale are sensitive only to $\mathcal{H}_{\rm IR}$, while the ultraviolet modes are effectively unobserved. The appropriate description of the proton at finite resolution is therefore given by the reduced density matrix
\begin{equation}
\rho_{\rm IR} = \mathrm{Tr}_{\rm UV} \, |p\rangle \langle p| .
\end{equation}

Even though the full state $|p\rangle$ is pure, the reduced density matrix $\rho_{\rm IR}$ is generically mixed. This reflects the entanglement between infrared and ultraviolet modes generated by QCD interactions. The corresponding von Neumann entropy,
\begin{equation}
S_{\rm IR} = - \mathrm{Tr} \left( \rho_{\rm IR} \ln \rho_{\rm IR} \right),
\end{equation}
is therefore nonzero and quantifies the information loss associated with restricting the description to a finite resolution scale.

The structure of the reduced density matrix can be formally expressed in terms of the modular (entanglement) Hamiltonian $K$,
\begin{equation}
\rho_{\rm IR} = e^{-K}.
\end{equation}
In general, $K$ is a nonlocal and complicated operator, and the reduced state is not thermal. However, in systems with strong interactions and significant mode mixing, the modular Hamiltonian can under certain conditions be approximated by a local operator proportional to the physical Hamiltonian~\cite{Bisognano:1976za, Swingle:2016foj} (for a review, see Ref.~\cite{Casini:2011kv}).  In this case, the reduced density matrix takes an approximately thermal form (see Ref.~\cite{Kharzeev:2026jkq} for a review): 
\begin{equation}
\rho_{\rm IR} \simeq \frac{1}{Z} \exp\left(-\frac{H_{\rm eff}}{T_{\rm eff}}\right),
\end{equation}
where $H_{\rm eff}$ is an effective Hamiltonian acting on the resolved degrees of freedom, and $T_{\rm eff}$ is an emergent temperature.

The emergence of such an effective thermal description is not automatic; it requires that the modular Hamiltonian be dominated by operators that are approximately local and energy-like. This condition is likely realized in QCD at the relevant scales, where strong interactions and multiparticle dynamics generate substantial entanglement between modes. In this regime, the reduced density matrix of the proton at finite resolution can be well approximated by a thermal ensemble.

In this sense, the proton acquires thermodynamic properties—such as entropy and temperature—that do not arise from coupling to an external heat bath, but instead emerge from quantum entanglement between resolved and unresolved degrees of freedom. This observation provides the conceptual basis for interpreting the entropy associated with the proton in terms of an effective thermal description, and for relating it to the entropy density of hadronic matter at the QCD transition scale.

\section{Thermodynamic Entropy of QGP --- and Protons}
\label{sec:thermodynamics}

With the motivations described in the Introduction,
let us start by observing that if we think of a proton as formed from a volume $V_p$ of QGP with temperature $T$, energy density $\varepsilon_{\rm QGP}$ and entropy density $s_{\rm QGP}$, then the mass of the proton is 
\begin{equation}
    M_p=V_p\,\varepsilon_{\rm QGP}
\end{equation}
and the thermal entropy of the QGP from which the proton forms is 
\begin{equation}
    S_{\rm thermal} = V_p\, s_{\rm QGP} = M_p \frac{s_{\rm QGP}}{\varepsilon_{\rm QGP}}\ .
\end{equation}
From lattice QCD calculations of the thermodynamics
of QGP~\cite{Borsanyi:2013bia,HotQCD:2014kol}, we see that $T s_{\rm QGP}/\varepsilon_{\rm QGP} \sim (1.1-1.2)$ over a range of temperatures around and above $T_c$, meaning that 
\begin{equation}
    S_{\rm thermal} \sim (1.1-1.2)\frac{938~{\rm MeV}}{T}\ .\label{eq:thermal-entropy}
\end{equation}
To turn this argument into an estimate of $S_{\rm thermal}/k_B$, we need to specify the temperature $T$ arising in Eq.~\eqref{eq:thermal-entropy}.
One obvious possibility is to take $T=T_c\sim (155-160)$~MeV~\cite{Borsanyi:2013bia,HotQCD:2014kol,HotQCD:2018pds,Borsanyi:2020fev,Borsanyi:2025dyp}, the temperature of the QCD crossover transition. This could be an underestimate, however, since as a protons form from a small droplet of QGP at the QCD phase transition, confinement forces can squeeze the droplet, which would increase both its internal pressure and the corresponding internal temperature $T_p$.
This suggests using what we know about the pressure within a proton to make an estimate.

Recent lattice QCD calculations of the gravitational form factors of the proton~\cite{Shanahan:2018pib,Shanahan:2018nnv,Hackett:2023rif} together with data from electron-scattering experiments~\cite{Polyakov:2002yz,Burkert:2018bqq,Lorce:2018egm,Polyakov:2018zvc} 
can be interpreted in terms of a pressure distribution $P(r)$ within a proton that depends on the radial distance $r$ from the center of a proton, with $r^2 P(r)\sim (0.002-0.004)~{\rm GeV}^2$
over a range of radii $\sim (0.15-0.4)$~fm~\cite{Hackett:2023rif}, 
corresponding to a pressure $P\sim (0.0005-0.005)~{\rm GeV}^4$ within a proton.
Lattice QCD calculations of $3P_{\rm QGP}/T^4$ for QGP~\cite{Borsanyi:2013bia,HotQCD:2014kol}
indicate that at $T=155/180/200$~MeV the pressure of QGP is  $\sim 0.0004/0.001/0.003$~GeV$^4$.  
It is pleasingly consistent that the pressure inside a proton is comparable to that of QGP with a temperature $T=T_p\sim (155-200)$~MeV that,
although we cannot determine in quantitatively, turns out to be comparable to or somewhat above $T_c$.
This suggests that in estimating $S_{\rm thermal}$ we may consider using 
either $T=T_c\sim (155-160)$~MeV or $T=T_p\sim(155-200)$~MeV in Eq.~\eqref{eq:thermal-entropy}.

Further support for thinking of a proton as if it had an internal temperature somewhere in this range 
comes via observing that if we think of a proton 
as made from a spherical droplet of QGP with volume
$V_p=(938~{\rm MeV})/\varepsilon(T_p)$ and temperature $T_p$ then 
the lattice QCD calculations of $\varepsilon/T^4$~~\cite{Borsanyi:2013bia,HotQCD:2014kol}
indicate that if we take $T_p\sim 155$~MeV we have a sphere with radius $\sim 0.86$~fm --- which is comparable to the charge radius of the proton --- whereas if we take $T_p \sim 200$~MeV we have a sphere with radius $\sim 0.49$~fm --- which is comparable to the mass radius of the proton~\cite{Kharzeev:2021qkd,Duran:2022xag}.

If we estimate $S_{\rm thermal}$ by setting $T=T_c=(155-160)$~MeV on the right-hand side of Eq.~\eqref{eq:thermal-entropy} based upon the idea that protons form from QGP at temperature $T_c$,
we find $6.5<S_{\rm thermal}/k_B<7.3$.
Or we can set $T=T_p\sim (155-200)$~MeV, based upon the above discussion of the internal pressure, energy density, and volume of the proton,
which yields $5.2<S_{\rm thermal}/k_B<7.3$.
To be conservative, the estimate with which we conclude this Section is that if we take seriously that a proton forms from QGP, 
according to the second law of thermodynamics 
the entropy within a proton should lie within or above the range
\begin{equation}
S_{\rm thermal}\sim (5-8)k_B\ .
\label{eq:S-thermal-estimate}
\end{equation}
This thermodynamic entropy must be converted 
into entanglement entropy of the quantum correlations among the confined partons in the wave function of the proton.

Later, in Section~\ref{sec:entanglement-entropy-via-Hagedorn}, we shall present an argument favoring an ``internal temperature'' of the proton $T_p$ that is less than the Hagedorn temperature $T_H\sim 167$~MeV, 
which favors the upper half of the range \eqref{eq:S-thermal-estimate} for $S_{\rm thermal}$.

\subsection{QGP Gibbs entropy as entanglement entropy}

Before turning to estimating the internal entanglement entropy of the proton on its own terms, we close this Section by noting that the Gibbs entropy $s_{\rm QGP} V$ of 
a volume $V$ of QGP can itself be understood as entanglement entropy. This argument is
more straightforward for the QGP produced in a heavy ion collision (a ``little Bang'') than for the Big Bang. Since the incident nuclei in a heavy ion collision are pure quantum states, and since evolution in quantum field theory is unitary, as a matter of principle the state
$|\Psi_{\mathrm{QCD}}(\tau)\rangle$ 
of the expanding fireball of matter a fixed 
proper time $\tau$ after the collision is a pure 
quantum state.
We then consider a sub-volume $V$ of QGP within the debris from the collision. The reduced density matrix for the QGP in the volume $V$ is
\begin{equation}
\rho_V = \mathrm{Tr}_{\overline{V}}
\big( |\Psi_{\mathrm{QCD}}(\tau)\rangle
      \langle \Psi_{\mathrm{QCD}}(\tau)| \big).
\end{equation}
According to the Maximal Entanglement Limit (MEL) conjecture (see Ref.~\cite{Kharzeev:2026jkq} and references therein),  $\rho_V$ is well approximated by a thermal density 
matrix with temperature $T(\tau)$ and baryon chemical potential
$\mu_B(\tau)$,
\begin{equation}
\rho_V \simeq
\frac{\exp\big[-\beta(\tau) H_V + \mu_B(\tau) B_V\big]}{
      \mathrm{Tr} \exp\big[-\beta(\tau) H_V + \mu_B(\tau) B_V\big]}.
\end{equation}
The von Neumann entropy of $\rho_V$,
\begin{equation}
S_{\rm ent}^{\rm QGP}(V)
= - \mathrm{Tr}\, \rho_V \ln \rho_V,
\end{equation}
is therefore equal, up to subleading corrections, to the Gibbs
entropy of the QGP in $V$:
\begin{equation}
S_{\rm ent}^{\rm QGP}(V)
\simeq S_{\rm Gibbs}^{\rm QGP}(V)
= s_{\rm QGP}(T,\mu_B)\, V,
\label{eq:GibbsQGP}
\end{equation}
with $s_{\rm QGP}$ the entropy density extracted from lattice QCD.
In this sense, the entropy of the volume $V$ of QGP
is exactly the
\emph{entanglement entropy between $V$ and its environment}.
The equivalence between the thermodynamic, Gibbs, entropy of equilibrated matter in a volume $V$ 
and the entanglement entropy between $V$ and its environment is a general result that is familiar in other contexts~\cite{Nandkishore:2014bkw,Abanin:2018yrt,Pilatowsky-Cameo:2024jkd}, that we have applied here to the Gibbs entropy of QGP.

\section{An Estimate of the Internal Entanglement Entropy of the Proton}
\label{sec:entanglement}

The observation that the thermal entropy of QGP can itself be viewed as entanglement entropy
gives us some confidence as we turn to investigating 
how the 
(large) Gibbs entropy carried by the deconfined QGP is encoded, after confinement, in a gas
of color-singlet hadrons.
We shall formulate this problem in the language of quantum
information theory and 
compare the entanglement entropy within a proton to the thermal entropy of the QGP from which that proton formed, Eq.~\eqref{eq:S-thermal-estimate}.
This viewpoint reveals a deep connection between the entropy of the
QGP and the entropy stored in the internal structure of hadrons as entanglement entropy.
To match the entropy carried by the deconfined quark--gluon plasma to the
entropy stored inside hadrons, we need to obtain a quantitative
estimate of the entanglement entropy inside a single proton.  
In this Section we provide one  estimate of this quantity using
 the known internal
structure of the proton in QCD.

First, in subsection \ref{defin} we will define what we mean by the internal entanglement entropy of the proton. Then in subsection \ref{hilb} we will estimate the effective dimension of the Hilbert space of a valence quark based on its discrete quantum numbers: color, spin, and flavor. Then, in subsection \ref{glu} we will argue that the number of gluonic degrees of freedom is large, which allows to apply Page's theorem to argue that the entanglement entropy of a quark with the rest of the proton is near the maximal allowed by the number of quark degrees of freedom. Finally, in subsection \ref{total} we will estimate the total entanglement entropy in the three-valence-quark system taking into account the color-singlet constraint and Fermi statistics.   

\subsection{Definition of the internal entanglement entropy}\label{defin}

Consider the proton in its rest frame.  
Let $|p\rangle$ denote its ground-state wavefunction in QCD.  
The Hilbert space factorizes as
\begin{equation}
\mathcal{H}
= \mathcal{H}_{q} \otimes \mathcal{H}_{\rm rest},
\end{equation}
where $\mathcal{H}_{q}$ is the local Hilbert space of a ``tagged''
quark (e.g.\ a valence up quark), and $\mathcal{H}_{\rm rest}$ contains
the remaining quarks, antiquarks, and gluons.  
The reduced density matrix of the tagged quark is
\begin{equation}
\rho_q = \mathrm{Tr}_{\rm rest} \big( |p\rangle\langle p| \big),
\end{equation}
and the single-quark entanglement entropy is
\begin{equation}
S_q^{(p)}
= - \mathrm{Tr} \,\rho_q \ln \rho_q.
\end{equation}
However, the total internal entanglement entropy  of the proton is not the
sum of $S_q^{(p)}$ for three quarks (because of correlations and
subadditivity); we will estimate it  below in subsection \ref{total}.  

\subsection{Local Hilbert space of a quark inside the proton}\label{hilb}

The effective dimension $d_q$ of $\mathcal{H}_{q}$ is determined by the
degrees of freedom available to a quark inside a confined hadronic
bound state in QCD:

\paragraph{Color:} $N_c = 3$.
\smallskip

\paragraph{Spin:} $2$ spin projections.
\smallskip

\paragraph{Flavor:} A ``tagged quark'' is not strictly flavor-fixed
because of sea-quark fluctuations.  The light flavors 
$(u,d,s)$ should be included; heavier flavors contribute negligibly.
Thus $N_f^{\rm eff} \simeq 3$.
\smallskip

\paragraph{Orbital and radial levels:}
The internal wavefunction of the proton includes at least the
lowest S-wave and a small admixture of excited modes; even a minimal
truncation corresponds to an effective factor that must be at least $\sim 2$.  However, we have no argument why this factor could not be larger than $\sim 2$, and our incomplete understanding of this factor is a dominant uncertainty in the estimate that we shall arrive at in this Section.

Thus, we take
\begin{equation}
d_q^{\rm(eff)} 
\gtrsim 3_{\rm color} \times 2_{\rm spin}
\times 3_{\rm flavor} \times 2_{\rm orbit}
\simeq 36\ ,
\label{eq:dq36}
\end{equation}
with the $\gtrsim$ arising because $2_{\rm orbit}$ could be an underestimate.
Because the quarks are confined inside the (small) volume of a proton, there are no continuous momenta to consider.

\subsection{Gluonic contributions, and  the Page theorem}\label{glu}

The gluon field contributes significantly to the internal structure of the
proton and to the entanglement of a tagged quark with the rest.  To understand the effect of gluons on the entanglement entropy, it is useful to invoke the Page theorem \cite{Page:1993}.

For a bipartite Hilbert space 
${\cal H} = {\cal H}_A \otimes {\cal H}_B$ with
$\dim {\cal H}_A = m \le n = \dim {\cal H}_B$, the average von Neumann
entropy of the subsystem $A$ for a random pure state on ${\cal H}$ is
\cite{Page:1993}
\begin{equation}
\langle S_A \rangle
= \sum_{k=n+1}^{mn} \frac{1}{k}
  - \frac{m-1}{2n}
\simeq \ln m - \frac{m}{2n} + \mathcal{O}\!\left(\frac{1}{n^2}\right),
\label{eq:PageFormula}
\end{equation}
where the last expression holds for $n \gg m$.

In our case, for the single-quark partition we have 
$m = d_q^{\rm(eff)} \gtrsim 36$ from Eq.~(\ref{eq:dq36}), and the
environment consists of the remaining two quarks plus the gluonic and
sea degrees of freedom.  A conservative estimate of the environment
dimension (which consists of two remaining valence quarks and gluons) is
\begin{equation}
n \sim (d_q^{\rm(eff)})^2 \times d_g,
\end{equation}
where $d_g$ is the effective dimension associated with the low-lying
gluonic and sea-quark modes in the proton.  
Even taking $d_g$ as small as $\mathcal{O}(10)$, 
which is likely a significant underestimate, we obtain
\begin{equation}
n \gtrsim 36^2 \times 10 \sim 1.3\times 10^4 \gg m.
\end{equation}
Inserting this estimate into Eq.~(\ref{eq:PageFormula}) and restoring the factor of $k_B$ yields
\begin{equation}
\frac{\langle S_q^{(p)} \rangle}{k_B}
\gtrsim \ln 36 - \frac{36}{2 n}
\simeq 3.58\ 
- \mathcal{O}(10^{-3})\ .
\end{equation}
We see that the correction to the maximal value $\ln d_q^{\rm(eff)}$ is at
the per-mille level or smaller.  In other words, once gluonic and sea
degrees of freedom are included, the tagged quark is essentially
\emph{maximally entangled} with the rest of the proton:
\begin{equation}
S_q^{(p)} 
\gtrsim 3.58\ k_B.
\label{eq:SqFinal}
\end{equation}
Thus the role of gluons here is not to change the leading value
$\ln d_q^{\rm(eff)}$ but to guarantee that the environment dimension
$n$ is large enough for Page's theorem to drive $S_q^{(p)}$ very close
to its maximal possible value.  The logic of this argument makes it clear that the single-quark entanglement entropy \eqref{eq:SqFinal}
is robustly determined by
$d_q^{\rm(eff)}$. 

\subsection{Total internal proton entropy}\label{total}

We now turn to the total internal entanglement entropy of the proton,
$S_{\rm entanglement}^{(p)}$.  To estimate it, we consider the three-valence-quark sector as our subsystem, and trace over
gluon and sea quark degrees of freedom. Via the Page theorem as above, the result will be determined by the dimension of the three-valence-quark Hilbert space.

For a single quark, we found $d_q^{\rm(eff)} \gtrsim  36$ including
color, spin, flavor, and a small number of orbital modes.  For three
quarks, the naive product space would have dimension
$(d_q^{\rm(eff)})^3 \gtrsim 36^3 = 46{,}656$.  However, physical three-quark
states are subject to the color-singlet constraint and Fermi statistics.

The color of three quarks transforms as
\begin{equation}
3\otimes 3 \otimes 3
= 1 \oplus 8 \oplus 8 \oplus 10,
\end{equation}
so only a one-dimensional singlet subspace is allowed.  This effectively
removes a factor of $3^3 = 27$ from the naive color count.  The relevant
three-quark Hilbert space for spin--flavor--orbital degrees of freedom
alone is then of order
\begin{equation}
d_{3q} \gtrsim (2_{\rm spin}\times 3_{\rm flavor}\times 2_{\rm orbit})^3
= 12^3 = 1{,}728.
\label{eq:d3q}
\end{equation}
Antisymmetrization further restricts this space, but it does not change
the order of magnitude of its dimension.

We now regard the three-quark color-singlet sector as subsystem $A$,
with dimension $m = d_{3q} \gtrsim 1.7\times 10^3$, and the gluons plus
sea quarks as environment $B$, with dimension $n \gg m$.  The Page formula
(\ref{eq:PageFormula}) then implies that the entanglement entropy between
the three-quark subsystem and the gluonic/sea environment is
\begin{equation}
S_{\rm entanglement}^{(p)} 
\simeq \ln d_{3q} - \frac{d_{3q}}{2 n}.
\end{equation}
Even for a very conservative environment dimension $n \gtrsim 10^5$,
the correction term $d_{3q}/(2n) \lesssim 0.01$ is negligible.  
We therefore obtain an estimate 
\begin{equation}
S_{\rm entanglement}^{(p)}
\gtrsim k_B \ln 1{,}728
\simeq 7.5.
\label{eq:SintTight}
\end{equation}

It is important to emphasize that Eq.~(\ref{eq:SintTight}) is subject
to model uncertainties in $d_{3q}$ (e.g.\ the number of effective
orbital modes included, detailed antisymmetrization, etc.), but these
uncertainties modify $\ln d_{3q}$ only at the level of a few tens of
percent in $d_{3q}$, i.e.\ by $\Delta S_{\rm entanglement} \sim \mathcal{O}( k_B)$.
By contrast, the Page-theorem correction $\sim d_{3q}/(2n)$ is at the
percent or per-mille level for any reasonable environment dimension.

To summarize, our estimate for the total entanglement entropy inside the proton  is
\begin{equation}
S_{\rm entanglement}^{(p)}
\gtrsim (7-8) k_B.
\label{eq:SintBenchmark}
\end{equation}
The estimate \eqref{eq:SintBenchmark} is fully consistent with the 
rougher estimate that we made in Section~\ref{sec:introduction} based on extrapolating from
previous calculations of the longitudinal component of the entanglement entropy
in Refs.~\cite{Hentschinski:2024gaa,Kharzeev:2021yyf}.
We note, however, that the most significant uncertainty in the
estimate \eqref{eq:SintBenchmark}, namely why 
truncate the orbital factor in the estimate \eqref{eq:dq36} at 2 rather than including higher energy orbital and radial levels, 
is completely different than 
the uncertainties in Refs.~\cite{Hentschinski:2024gaa,Kharzeev:2021yyf}.

Most striking,
our estimate \eqref{eq:SintBenchmark} for
the total entanglement entropy inside a proton
is remarkably similar to the estimate \eqref{eq:S-thermal-estimate} for the Gibbs entropy of the QGP from which each proton formed
at the time of the QCD phase transition, around ten microseconds after the Big Bang.
The comparison between the estimates \eqref{eq:S-thermal-estimate} and \eqref{eq:SintBenchmark} is beautifully consistent with the second law of thermodynamics, applied to the process via which a proton forms from QGP,  which dictates
that $S^{(p)}_{\rm entanglement}$ must be greater than or equal to $S_{\rm thermal}$.

\subsection{Internal entanglement entropy of mesons}

A similar, though quantitatively smaller, internal entanglement entropy can be
associated with mesons.
A meson consists of a quark--antiquark pair accompanied by gluonic degrees of freedom,
with a smaller internal Hilbert space than that of baryons.
Repeating the reduced-density-matrix analysis for mesons, one finds
\begin{equation}
S_{\rm entanglement}^{({\rm meson})}
\sim
\ln \dim \mathcal{H}^{\rm int}_{M}
\gtrsim \mathcal{O}(3\text{--}5),
\label{eq:meson_entropy_estimate}
\end{equation}
depending on the number of active gluonic modes at the resolution scale of interest.
Compared to baryons, mesons have fewer color-carrying constituents and lack a baryon
junction structure, leading naturally to a smaller entanglement entropy.

\section{Internal Entanglement Entropy of Protons from the Hagedorn Spectrum}
\label{sec:entanglement-entropy-via-Hagedorn}

In making the two estimates of the entanglement entropy within a proton that we have made in 
Sections~\ref{sec:introduction}
and \ref{sec:entanglement}, we have in different ways begun from considerations of what we know about the ground state of the proton.  In this Section, we take a different approach to making such an estimate, where we begin  from the observation of Section~\ref{sec:thermodynamics} that the interior of a proton has
the same pressure as QGP with some temperature $T_p$. Here we ask how we would estimate the internal entanglement entropy of the proton if we attempt to take the notion of an internal effective temperature $T_p$ seriously.
We defer to the next Section a somewhat formal demonstration 
that even though the proton is a pure state if the density matrix that describes the proton has the feature that after tracing over unobserved degrees of freedom the reduced density matrix looks thermal, it can make sense to assign an effective internal temperature to the proton.  In this Section, we pursue the consequences of this idea --- which implies that the probability that a proton is found in an excited state with mass $M>M_p$ is proportional 
to $\exp[-(M-M_p)/T_p]$ --- and obtain a third estimate of the
internal entropy of the proton.

To pursue the consequences of this speculation, we need the density of excited resonance states of the proton as a function of $M$.
Famously, the Hagedorn density of 
all resonance
states, that we shall denote by $\rho^H(M)$, grows exponentially at large $M$, with this growth serving to define the Hagedorn temperature $T_H$ via
$\rho^H(M)\propto \exp(M/T_H)$~\cite{Hagedorn:1965st,Hagedorn:1968zz}. 
We shall employ the contemporary determination of $\rho(M)$ from Ref.~\cite{SanMartin:2023zhv} based upon the PDG2021+ list of more than 3000 hadrons:
\begin{equation}\label{eq:rho-form-full-HRG}
\rho^H(M)=\frac{A}{\left(M^2+M_r^2\right)^{5/4}}\exp\left(\frac{M}{T_H}\right)\ ,
\end{equation}
where $M_r=500$~MeV and where the values of the parameters $T_H=167.2$~MeV and $A=0.4735$~GeV$^{3/2}$ were obtained in Ref.~\cite{SanMartin:2023zhv} via fitting to the density of states in the PDG2021+ list of hadrons. 
What we need, however,
is the density of states for only those hadronic resonances that have the same quantum numbers as the proton that we shall denote by $\rho_p^H(M)$. It is reasonable to assume that this takes the form
\begin{equation}\label{eq:rho-form-proton-HRG}
\rho^H_p(M)=\frac{A_p}{\left(M^2+M_r^2\right)^{5/4}}\exp\left(\frac{M}{T_H}\right)\ .
\end{equation}
It has long been understood that $T_H$ for baryonic resonances alone has the same value as $T_H$ for mesonic resonances alone,
see for example Ref.~\cite{Cleymans:2011fx}.  
This can be understood as arising from the relationship between $T_H$ and the QCD string tension, noting that the same string tension describes the string between a quark and antiquark in a meson and the strings between a baryon junction and the quarks in a baryon~\cite{Rossi:1977cy,Kharzeev:1996sq,Frenklakh:2024mgu}. We shall therefore take $T_H=167.2$~MeV in Eq.~\eqref{eq:rho-form-proton-HRG} as in Eq.~\eqref{eq:rho-form-full-HRG}.  
Although it is clear that $A_p$ must be less than $A$, we do not know its value. We shall see below, though, that our results depend on $A_p$ only logarithmically.

If we take $T_p$ seriously, and think of it as an effective internal temperature within a proton, this allows us to
write the probability $p(M)$ that a proton is found in one of its excited states with mass $M$ as
\begin{equation}\label{eq:p-defn}
    p(M)\equiv \frac{1}{Z} \rho^H_p(M) \exp\left(-\frac{M-M_p}{T_p}\right)
\end{equation}
where 
\begin{equation}\label{eq:Zdefn}
Z\equiv \int_{M_p}^\infty dM \rho_p(M)\exp\left(-\frac{M-M_p}{T_p}\right)\ ,
\end{equation}
which is finite, with the Boltzmann factor playing the role of an ultraviolet cutoff,
as long as $T_p<T_H$ --- which we shall assume henceforth. We note that since $Z$ and $\rho^H_p$ are proportional to $A_p$, the probability distribution $p(M)$ is independent of $A_p$.  For the same reason,
$p(M)$ also does not depend on the factor $\exp[+M_p/T_p]$ arising in its definition.
We can then estimate the internal entanglement entropy of the proton as the Boltzmann (or Shannon) entropy of the ensemble described by this probability distribution:
\begin{eqnarray}\label{eq:entanglement-entropy-from-p}
\frac{S_{\rm entanglement}^{\rm (p)}}{k_B} &=& - \int_{M_p}^\infty dM \rho_p^H(M) \frac{p(M)}{\rho^H_p(M)} \ln \left( \frac{p(M)}{\rho^H_p(M)} \right)\nonumber\\ 
&=& - \int_{M_p}^\infty dM  p(M) \ln \left( \frac{p(M)}{\rho^H_p(M)} \right)\ ,
    \end{eqnarray}
where $p(M)/\rho^H_p(M)$ is the probability (not probability density) that the proton is found in 
the excited state within $1/\rho^H_p(M)$ of $M$.
We see that, as we noted above, $S^{(p)}_{\rm entanglement}$ depends on $A_p$ only logarithmically
since $p(M)$ is independent of $A_p$. 
The expression \eqref{eq:entanglement-entropy-from-p}
suffices for our purposes here, but 
it will be helpful in the next Section 
to note that we can use Eq.~\eqref{eq:p-defn}
to recast  Eq.~\eqref{eq:entanglement-entropy-from-p} as
\begin{eqnarray}\label{eq:entanglement-entropy-from-Z}
\frac{S^{(p)}_{\rm entanglement}}{k_B} &=& \int_{M_p}^\infty dM p(M) \left(\frac{M-M_p}{T_p} +\ln Z\right)\nonumber\\ 
&=& \frac{\langle M-M_p\rangle}{T_p} + \ln Z \nonumber\\
&=& \frac{\langle M \rangle}{T_p} + \ln \tilde Z\ ,
\end{eqnarray}
where the expectation value is defined via the probability distribution $p(M)$
and where 
\begin{equation}\label{eq:Ztilde-defn}
\tilde Z \equiv 
    \int_{M_p}^\infty dM \rho_p(M)\exp\left(-\frac{M}{T_p}\right)\ .
\end{equation}

We can use either Eq.~\eqref{eq:entanglement-entropy-from-p} or Eq.~\eqref{eq:entanglement-entropy-from-Z} 
to make explicit estimates.
If we set $A_p=A$ and $T_H=167.2$~MeV, then for $T_p=155/160/165$~MeV 
we obtain the estimates $S_{\rm entanglement}^{\rm (p)}/k_B\simeq 7.6/8.6/10.8$.
With the more plausible choice $A_p=A/10$ and $T_H=167.2$~MeV, we obtain
$S_{\rm entanglement}^{\rm (p)}/k_B\simeq 5.3/6.3/8.5$, again for $T_p=155/160/165$~MeV.
In this way, we arrive at our third estimate of the internal entanglement entropy of the proton:
\begin{equation}\label{eq:SintBenchmark2}
    S^{(p)}_{\rm entanglement}\sim (5-9)k_B\ ,
\end{equation}
where here the dominant uncertainty comes from our lack of knowledge of $\ln(A_p/A)$, namely the logarithm of the ratio of the Hagedorn density of states in a hadron gas made up only of hadrons with the same quantum numbers as the proton to that in the conventional hadron resonance gas.
We also note that this estimate relied upon taking the notion of an internal temperature of the proton seriously. We investigate this idea further in the next Section.

Our  estimates~\eqref{eq:SintBenchmark} and \eqref{eq:SintBenchmark2} for
the total entanglement entropy inside a proton,
with very different dominant uncertainties,
are pleasingly similar to each other as well as to the more crude estimate that we made in Section~\ref{sec:introduction}.
And, as already noted, all of these estimates are strikingly similar to the estimate \eqref{eq:S-thermal-estimate} for the Gibbs entropy of the QGP from which each proton formed at the time of the QCD phase transition 
and thus consistent with the second law of thermodynamics as applied to the formation of protons from QGP.

\section{Effective internal temperature of the proton}
\label{sec:internal-temperature}

We have come to the notion of an internal temperature of the proton $T_p$ that is close to or perhaps slightly above the QCD crossover temperature $T_c$ twice, from two quite different perspectives.  In Section~\ref{sec:thermodynamics},
we came to this via realizing that when the first protons formed in the early universe they formed from QGP with $T\sim T_c$ and that the internal pressure within protons corresponds to that of QGP with 
a temperature $T_p\gtrsim T_c$. 
Then, in Section~\ref{sec:entanglement-entropy-via-Hagedorn} we discovered that if we take $T_p$ 
seriously as an internal temperature then, as long as $T_p<T_H\sim 167$~MeV, it serves to render the probability $p(M)$ that a proton is found 
in one of its excited states with mass $M$ finite.
Both perspectives led to estimates of the internal (entanglement) entropy of the proton, estimates that proved to be strikingly similar.

In this Section we show in a simple, but somewhat formal, fashion how to assign an effective internal temperature to a proton --- which is after all a pure state.
Let \(\rho_p\) denote the density matrix describing the internal
quantum state of the proton,\footnote{The Greek letter $\rho$ is conventionally used for both the Hagedorn density of states and for density matrices --- which are completely different objects. We use $\rho_p$ for the density matrix describing the measured subspace of the quantum state of a proton and $\rho^H_p(M)$ for the Hagedorn density of excited states of the proton with mass $M$.} 
and let $H$ be the QCD Hamiltonian.
The total internal energy of a proton can then be writen as
\begin{equation}
    E_p
    =
    {\rm Tr}\,\rho_p\,H .
\end{equation}
The essential physical assumption
that leads us directly to an effective internal temperature for the proton
is that, after tracing over degrees of
freedom that are not observed, the reduced density matrix of the measured
subsystem can be approximated by a thermal density matrix:
\begin{equation}\label{eq:thermal-density-matrix}
    \rho_p
    =
    \frac{1}{Z(\beta)}\,e^{-\beta H},
    \qquad
    Z(\beta)
    =
    {\rm Tr}\,e^{-\beta H},
\end{equation}
with
\begin{equation}
    \beta \equiv \frac{1}{ T_{\rm eff}} .
\end{equation}
Here \(T_{\rm eff}\) is not the temperature of an externally equilibrated heat
bath.  Rather, it is some effective internal temperature characterizing the
entanglement spectrum of the pure quantum mechanical state of the proton, that to this point
we have not specified.

The value of \(T_{\rm eff}\) is fixed by matching the expectation value of the
Hamiltonian to the proton energy,
\begin{equation}\label{eq:Ep}
    E_p
    =
    {\rm Tr}\,\rho_p H
    =
    -\frac{\partial}{ \partial \beta}\ln Z(\beta).
\end{equation}
In the proton rest frame one may take
\begin{equation}
    E_p \simeq M_p ,
\end{equation}
or, more generally, replace \(M_p\) by the invariant mass \(M_{\rm eff}\) of the
subsystem probed in a given process.  
Requiring that when the thermal density matrix $\rho_p$ from Eq.~\eqref{eq:thermal-density-matrix}
is substituted into the expression \eqref{eq:Ep} it yields $E_p\simeq M_p$ serves to fix the value of $T_{\rm eff}$.\footnote{We thank Mark Srednicki for pointing out this argument, and for noting that it is analogous to the 
eigenstate thermalization hypothesis~\cite{Deutsch:1991msp,Srednicki:1994mfb}.} 

We can pursue this argument further, by evaluating the von Neumann entropy corresponding to 
the density matrix of the measured subspace of the quantum mechanical state of the proton, assumed thermal:
\begin{equation}
    S_{\rm vN}/k_B
    \equiv
    -{\rm Tr}\,\rho_p\ln\rho_p \, .
\end{equation}
Using the thermal form of \(\rho_p\), one obtains
\begin{align}
    S_{\rm vN}/k_B
    &=
    -{\rm Tr}\,\rho_p
    \left(
        -\beta H-\ln Z
    \right)
    \nonumber\\
    &=
    \beta\,{\rm Tr}\,\rho_p H+\ln Z
    \nonumber\\
    &=
    \beta E_p+\ln Z(\beta).
\end{align}
Thus, the entanglement entropy associated with the thermal reduced density
matrix $\rho_p$ and the internal effective temperature $T_{\rm eff}$ is
\begin{equation}\label{eq:SvN}
    S_{\rm vN}/k_B
    =
    \frac{E_p}{ T_{\rm eff}}
    +
    \ln Z(T_{\rm eff}) .
\end{equation}
To this point, the argument of this Section may have seemed formal.  We now see, however, that 
if we take $T_{\rm eff}=T_p$ and assume that the 
$\ln Z$ in Eq.~\eqref{eq:SvN} is the same as the $ \ln \tilde Z $ in
Eq.~\eqref{eq:Ztilde-defn}, then the formal expression~\eqref{eq:SvN} agrees with
the expression~\eqref{eq:entanglement-entropy-from-Z}
for the entanglement entropy of the proton that we have evaluated explicitly in the previous Section.
In this way, the more formal argument of this Section supports the interpretation of the proton as a highly entangled quantum system whose reduced density matrix
is approximately thermal, characterized by an effective internal temperature $T_p$.

\section{Discussion and a Look Ahead}
\label{sec:conclusions}

We have presented four different estimates of the internal entropy of a proton. 
In Section~\ref{sec:thermodynamics},
we estimated the thermal Gibbs entropy of the QGP from which a proton forms at the QCD phase transition.  In Sections~\ref{sec:introduction},
\ref{sec:entanglement} and \ref{sec:entanglement-entropy-via-Hagedorn} we estimated the entanglement entropy within a proton, doing so
in Section~\ref{sec:introduction}
via a rough extrapolation from
the longitudinal entanglement entropy of one quark with the rest of the proton determined via deep inelastic scattering, 
in Section~\ref{sec:entanglement} via a counting argument starting from basic facts about the structure of a proton 
together with Page's theorem,
and in Section~\ref{sec:entanglement-entropy-via-Hagedorn} via 
taking the notion of an effective internal temperature of the proton seriously and using this notion together with the Hagedorn density of excited states of a proton.
All four of these estimates are qualitative, none is rigorous, and all have completely different conceptual and quantitative uncertainties.
And yet they all point to
similar values.

The
striking agreement between the different estimates of the entropy, thermodynamic or entanglement, within a proton serves at present more to pose provocative questions than
to provide definitive answers. The questions that occur to us include:
\begin{itemize}
    \item 
    How can we turn the idea that at the QCD crossover transition the thermodynamic entropy of a hadron-sized droplet of QGP is rearranged into quantum entanglement entropy within the hadron itself to better understand the process of hadronization in heavy ion collisions via freezeout of QGP and/or to improve phenomenological models of freezeout?
    \item 
    When experimentalists collide nuclei in top energy collisions at  RHIC or at the LHC, the QGP produced in the collisions has more entropy than the internal entanglement entropy within the incident nucleons, because the QGP produced is much hotter than $T_p$.  How can we learn about the liberation of entanglement entropy and the formation of thermodynamic entropy by focusing on lower energy heavy ion collisions that produce QGP with a temperature $\sim T_p$?
    \item 
    In a brutal sense, confinement is the answer to the question of why a proton-sized volume of matter with an effective internal temperature $T_p$ and a corresponding, large, internal pressure does not explode. The rearrangement of thermal entropy into entanglement entropy appears more subtle, less brutal. Can one be used to better understand the other?
    \item 
    Can measurements of hadronic entanglement --- for example, through correlations in final states of heavy ion collisions~\cite{Hutson:2024xsv} --- provide novel probes of the QCD phase transition?
    \item 
    Can extant lattice QCD studies that indicate an internal pressure within the proton~\cite{Shanahan:2018pib,Shanahan:2018nnv,Hackett:2023rif} 
    be related to future lattice~\cite{Amorosso:2024leg,Amorosso:2024glf,Mamo:2026ktr} or Hamiltonian lattice~\cite{Horn:2026bfs} calculations of entanglement entropy within hadrons? How do the results of such calculations depend on the lattice spacing?
    \item 
    We noted in Sections~\ref{sec:entanglement-entropy-via-Hagedorn} and \ref{sec:internal-temperature} that since a proton has a similar internal energy density and pressure to a droplet of QGP with the same mass and a temperature $T_p$, perhaps we should take $T_p$ seriously as an internal effective temperature.  This led us to one of our estimates of the internal entanglement entropy of the proton, which turned out to be strikingly similar to the Gibbs entropy of the droplet of QGP from which a proton forms at the QCD transition. What else can we learn about protons from what we know about QGP? QGP at temperatures $\sim T_p$ is a strongly coupled fluid, with strong correlations between the momenta of fluid cells that are near each other in position space. This suggests that there are strong correlations between the momenta of nearby quarks and gluons within a proton. How can such correlations be measured at the Electron-Ion Collider?
    \item It is reasonable to speculate that the same strong correlations are responsible for the entanglement entropy within a proton.  How can this be measured at the Electron-Ion Collider?
    \end{itemize}
We hope that sharing 
the qualitative observations
that we have made in the earlier Sections of this paper
is as thought-provoking for others as it is for us, and we hope that it prompts many
discussions, of the questions that we have enumerated and more. 
Of course we cannot rule out the possibility that the striking agreement between the four estimates of the internal entropy of a proton is a quadruple coincidence. We rather hope, though, that
probing this agreement and the questions that it raises  will
further advance our understanding of confinement,
hadron structure, quark-gluon plasma, and the QCD phase transition via all the ways in which entropy --- whether understood as thermal or quantum --- is woven throughout.

\medskip

\acknowledgments

This work was initiated in part at the Aspen Center for Physics, which is supported by National Science Foundation grant PHY-2210452. We gratefully acknowledge conversations with many participants at the Strongly Interacting Matter at the Electron-Ion Collider workshop
after one of us gave a seminar about entanglement entropy and the other one of us gave a seminar about thermodynamic entropy.
KR is also grateful to the Kavli Institute for Theoretical Physics (KITP) for hospitality and support as this work was completed and acknowledges conversations with many participants at the Frontiers of Quark-Gluon Matter Program; this research was supported in part by grant NSF PHY-2309135 to the KITP.

We particularly thank Soonwon Choi, Jamie Karthein,
Zein-Eddine Meziani, Richard Milner, Berndt M\"uller, Dimitra Pefkou, Jen-Chieh Peng, Felix Ringer, Mark Srednicki, John Terry, 
Xiaojun Yao and Yong Zhao for helpful conversations.

Research of KR was supported in part by the U.S.~Department of Energy, Office of Science, Office of Nuclear Physics under grant Contract Number DE-SC0011090. The work of DK was supported by the U.S. Department of Energy, Office
of Science, Office of Nuclear Physics under contract No.DE-FG88ER41450
and by the U.S. Department of Energy, Office of Science, National Quantum Information Science Research Centers, Co-design Center for Quantum
Advantage (C2QA) under Contract No.DE-SC0012704.

\bibliographystyle{unsrtnat}
\bibliography{main,main2}

\end{document}